\documentclass[11pt,a4paper]{article}

\usepackage[T1]{fontenc}
\usepackage[utf8]{inputenc}
\usepackage{lmodern}
\usepackage[english]{babel}
\usepackage[margin=2.5cm]{geometry}
\usepackage{microtype}
\usepackage{graphicx}
\usepackage{booktabs}
\usepackage{tabularx}
\usepackage{siunitx}
\usepackage{xcolor}
\usepackage{listings}
\usepackage{tikz}
\usetikzlibrary{positioning,arrows.meta,fit}
\usepackage[round]{natbib}
\usepackage[hidelinks]{hyperref}
\usepackage[capitalise,noabbrev]{cleveref}

\lstdefinestyle{tok}{basicstyle=\ttfamily\small,columns=fullflexible,keepspaces=true,frame=single,framesep=4pt,xleftmargin=6pt,xrightmargin=6pt,showstringspaces=false,breaklines=true}
\title{tok: a textual concrete syntax of BPMN and DMN for humans and AI agents}
\author{Marek Deja\\[2pt]\small Jagiellonian University, Kraków, Poland\\\small ORCID 0000-0003-0015-2322}
\date{30 October 2026}

\begin{document}

\maketitle

\begin{abstract}
tok is a plain-text notation for Business Process Model and Notation (BPMN) 2.0.2 process models and Decision Model and Notation (DMN) 1.5 decisions. One engine checks a tok document against both standards, compiles the document to XML with an automatic diagram, and serves language-model applications through the Model Context Protocol. The paper classifies the notation as an alternative concrete syntax of the BPMN and DMN metamodels and reports one study. A tok model takes 2.2 times fewer tokens than the smallest BPMN XML of 893 files exported by 37 tools and 7.7 times fewer than the files as exported. The checker signals 98.2\% of 1\,411 injected faults, against 85.8\% for the XML Schema with bpmnlint. Four language models new to tok write valid tok from a one-page reference for 85.7 to 100\% of 87 process descriptions from two datasets and valid BPMN XML for every description, and the strongest model writes tok with 25\% and 19\% fewer tokens than BPMN XML. As agents with a lean skill and one checker tool, the models write valid tok for 97.6 to 100\% of the descriptions, the strongest model spends 21\% and 14\% fewer tokens on tok than on BPMN XML, and a skill which sends the model to read the reference through a tool costs 1.7 to 2.8 times the tokens of the lean skill. Of 47 event logs simulated from public models, 44 fit their source models, and tok imports every model discovered from the logs.
\end{abstract}

\section{Introduction}\label{sec:intro}

Organisations keep models of their processes as information objects, which document work, support analysis and compliance, configure process engines and teach students how work is organised \citep{Dumas2018fundamentals}. Business Process Model and Notation (BPMN) 2.0.2 is the standard notation for process models, with an XML format for the exchange of models and their diagrams between tools \citep{OMG2014BPMN}, and Decision Model and Notation (DMN) 1.5 is the companion standard for the decisions taken inside processes \citep{OMG2024DMN}. Practice uses a small part of the BPMN vocabulary, as \citet{zurMuehlen2008} found that a few core constructs account for most of the elements in the models of their sample, and a survey of practitioners reports the uptake of the notation together with the difficulties of its users \citep{Recker2010opportunities}.

A process model has three readers with different needs. A person reads a diagram and changes the model in a tool, a process engine and an exchange partner read the XML file, and a language model, the newest reader, writes and reads whatever text the prompt contains. The exchange format serves the second reader, because a correct file needs unique identifiers, references in both directions between nodes and flows, the element order of the schema and, for a diagram, the coordinates of every shape and every bend of an edge. People therefore read, review and compare the files with difficulty, since reading a notation is a skill of its own \citep{Petre1995looking} and the comprehension of a process description depends on the notation and on the training of the reader \citep{Ottensooser2012making,Figl2017comprehension}, and a version-control difference of a small change to a drawn model consists mostly of geometry, a known cost of versioning graphical models \citep{Altmanninger2009survey}. The third reader meets the same cost, in that a model written from scratch in BPMN XML spends most of its output on syntax and references and valid BPMN XML remains a hurdle for open models \citep{Lauer2026competences}, and research on process modelling with language models therefore works through intermediate representations, for example process code in the POWL language translated to BPMN by a converter \citep{Kourani2024}, and evaluates conversational modelling and process tasks \citep{Klievtsova2023,Grohs2024}. The intermediate representations serve the model, while the person keeps the diagram and the file, so that the three readers of one process still hold three artefacts.

tok is a plain-text notation for the content of BPMN 2.0.2 and DMN 1.5, written for the three readers at once. A tok document states pools, lanes, events, activities, gateways, sequence and message flows, data, decisions, forms and simulation parameters in short lines, one statement per line, and one engine compiles the document to BPMN XML with an automatic diagram layout, SVG pictures, DMN files, Camunda 8 forms and simulated event logs. The engine reads BPMN and DMN files drawn in other tools and prints the files as tok, checks documents and BPMN files against the rules of both standards, and offers the same operations to language-model applications through a Model Context Protocol (MCP) server \citep{MCP2025}. The proposition behind the design is that one symbolic notation over the metamodel of the standard, with a checker which answers every text with a diagnostic at the line and the element concerned, serves as the shared code of a person, a tool and a language model, so that the document which the person keeps is the text which the model writes, and the file which the tool reads is compiled from both.

The paper makes four contributions, and the first contribution is the notation and its engine, described as a design which covers the syntax, the levels and conformance classes, the checker and the interfaces, and which shows how a textual notation stays exact against two standards. The second contribution is a classification of tok in formal language theory, model-driven engineering, semiotics and the theory of communication, which explains why the translation between tok and the standards is exact and why the diagnostics of the checker work as common ground between a writer and the engine. The third contribution is one study with six measurements, namely the size of tok against BPMN XML in the tokens that language models read, the detection of injected modelling faults, the writing of process models by four language models from two sets of texts, a process-mining demonstration, the speed of the engine and its conformance suites, which together put numbers on the price and the benefit of the notation for each of the three readers. The fourth contribution consists of the material of the study, which makes the software, the data and the scripts available for further work. The results support the proposition in measured terms, in that a tok model takes 2.2 times fewer tokens than the smallest BPMN XML of 893 tool files and 7.7 times fewer than the files as exported, the checker signals 98.2\,\% of 1\,411 injected faults against 85.8\,\% for the XML validators. Four language models which meet tok for the first time write valid tok from one page of reference for 85.7 to 100\,\% of 87 texts, with 25\,\% and 19\,\% fewer tokens than BPMN XML for the strongest model; as agents with a lean skill and one checker tool the models write valid tok for 97.6 to 100\,\% of the texts, and the strongest model spends 21\,\% and 14\,\% fewer tokens on tok than the same agent spends on BPMN XML; and 44 of 47 event logs simulated from public models fit their source models under replay.

\section{Background and related work}\label{sec:background}

\subsection{The standards and their exchange format}

BPMN 2.0.2 defines a metamodel of process models, a graphical notation, execution semantics for a part of the metamodel, and an XML format with a schema for the exchange of models, in which a separate diagram interchange part stores the position of every shape and every edge \citep{OMG2014BPMN}, and DMN 1.5 defines decision requirements graphs, decision tables and boxed expressions with the expression language FEEL, with an XML format of the same construction \citep{OMG2024DMN}. The separation of the model from the picture in both standards has a consequence for the design of any further notation, because a file without diagram interchange is a complete model under the standard, so that a notation may carry the model alone and leave the picture to a layout algorithm without any loss against the standard. tok rests on the separation, in that the text holds the model and the engine computes the drawing.

The exchange format serves tools, and tools disagree about the format. \citet{Geiger2018bpmn} found that modelling tools and engines differ in the parts of BPMN 2.0 which are supported and in the files which are written for the same model, and the Model Interchange Working Group collects the files that modelling tools export for a set of reference models in a public test suite, because interchange between tools fails often enough to need such a test \citep{MIWG}. A notation which claims an exact translation to the standard therefore has to be measured against the files that tools write, and the study takes 893 files of 37 tools from the suite for the purpose.

The vocabulary of BPMN is large and the use of the vocabulary is concentrated. \citet{zurMuehlen2008} found that a small set of constructs accounts for most of the elements in the models of their sample and that many constructs are rarely used, guidelines for modelling recommend small models with few elements \citep{Mendling2010}, and the formal semantics of BPMN through Petri nets shows where the informal semantics of the standard leaves cases open \citep{Dijkman2008}. The concentration sets two requirements for a notation at once, since a reference of one page can cover the core that writers use, whereas a notation which promises a round trip with tool files has to cover the whole metamodel, and tok meets both requirements with a one-page reference, which the study gives to the language models, and a vocabulary of language 1.0 which spans BPMN 2.0.2 and DMN 1.5.

\subsection{Textual notations for process models}

Textual notations for process models occupy three positions, which differ in what the text stands for. In diagram languages such as Mermaid and PlantUML the text stands for a picture, and BPMN Sketch Miner derives BPMN diagrams from a compact outline of activities and actors, which shows that a short text suffices for sketching a process and delivers the diagram as the product \citep{Ivanchikj2020}. In POWL the text is code for a partially ordered workflow, sound by construction and converted to BPMN or to Petri nets \citep{Kourani2023powl}, and in the model-driven engineering of \citet{Drave2022mde} a textual BPMN language with a soundness checker serves the generation of code for process-aware information systems. In BPCF, the third position, the text is a notation designed for language models to write, lossless for the constructs of the PMo benchmark and compared with Mermaid, JSON and BPMN XML in tokens, cost and accuracy across five language models \citep{Agrawal2026bpcf}.

tok takes a position defined by the metamodel of the standard. Among the design decisions for domain-specific languages, the reuse of an existing abstract syntax under a new concrete syntax keeps the semantics of the host and limits the new work to the mapping \citep{Mernik2005}, and tok applies the decision to the BPMN and DMN metamodels, so that the text covers both metamodels, the checker applies the rules of the standards, an imported tool file prints as tok and exports back, and the same text serves a person as a document. The position has a measurable price and a measurable benefit, which the study reports as the size of the text against the XML of tool files, the fidelity of the round trip, and the detection of faults by the checker.

\subsection{Language models and process models}

The generation of process models from text has moved from rules to language models, and the representation that the model writes has become a variable of the research. Rule-based pipelines produced BPMN models from text \citep{Friedrich2011}, the PET dataset supplied annotated descriptions for the evaluation of extraction methods \citep{Bellan2022EDOC,Bellan2023PET}, and large language models opened a direct route from text to model together with a list of open questions for process management \citep{Vidgof2023llm}. The direct route runs into the exchange format, since a benchmark of 17 open models found valid BPMN XML a major hurdle \citep{Lauer2026competences}, and the field answers with intermediate representations, such as POWL code with a converter and a loop of error feedback in ProMoAI \citep{Kourani2024,Kourani2026}, a JSON layer in BPMN Assistant \citep{Licardo2026assistant}, a Python-like representation compiled to BPMN and DMN in FLOW-BENCH \citep{Duesterwald2025flowbench}, formal steps after the model in the pipeline of \citet{Nivon2025generation}, and instruction tuning of open-weight models on pairs of text and model \citep{Celikmasat2025instruction}.

The choice of representation changes the outcome, and the measurements so far leave the shared document aside. \citet{Brissard2026representation} compared nine representations on 55 descriptions and found Mermaid best overall, a compact intermediate format cut the tokens of conversational modelling against ProMoAI \citep{Kopke2025efficient}, studies of conversational modelling measured the quality of models built in dialogue \citep{Klievtsova2023}, and \citet{Grohs2024} evaluated language models on the extraction of imperative and of declarative models and on the assessment of tasks for automation. Outside process modelling, the format of a prompt alone changes the performance of a model \citep{He2024formatting}. The representations in the studies named are built for the model, with the person reading a converted diagram, and the quality criteria differ between the studies, from the similarity of elements to the validity of the XML. The study of tok therefore judges both representations with one external judge, the OMG schema and the correctness rules of bpmnlint, on four models of three families and two datasets, and the representation under test is the document which the person keeps.

\subsection{Language models with tools}

The behaviour of a language model with tools depends on the interface of the tools and on the feedback which the tools return, and both are design choices of the environment. A model which interleaves reasoning with calls to tools corrects course from observations \citep{Yao2022ReAct}, a model can learn when and how to call a tool \citep{Schick2023}, surveys of tool learning catalogue the forms which the interface takes \citep{Qin2025tool}, feedback from execution lets a model repair its own output \citep{Chen2024selfdebug}, and in software engineering the design of the interface between the agent and the computer, with concise feedback and guard rails, decides what the same model achieves \citep{Yang2024sweagent}. The Model Context Protocol standardises how an application offers tools, resources and prompts to a model \citep{MCP2025}, and the ecosystem of servers raises questions of security and maintenance \citep{Hou2025mcp}. For a notation which a model meets for the first time, the findings turn the question of learnability into a question about the feedback channel, that is whether a page of reference with the messages of a checker suffices for valid models, and what a skill document and tools called by the model add, at what price in tokens. The study answers the question by treating the two channels as conditions of one design under the same judge.

\subsection{Simulation and process mining}

Process mining reads logs and writes models, and a notation that holds the model and its simulation parameters in one text can stand at both ends of the loop. Discovery derives a model from an event log and conformance checking replays a log on a model \citep{vanderAalst2016}, with XES as the standard format for logs \citep{IEEE1849}, and discovered models are judged by fitness, precision, generalisation and simplicity \citep{Buijs2014quality}, with a benchmark of discovery algorithms on public logs \citep{Augusto2019discovery}. The inductive miner discovers block-structured sound models which fit the log \citep{Leemans2013}, token-based replay measures fitness by replaying traces on a Petri net \citep{Berti2021tbr}, PM4Py implements discovery, conformance checking and the conversion between BPMN and Petri nets \citep{Berti2023PM4Py}, and business process simulation produces logs from models with durations, resources and branching probabilities, with simulation models discovered from logs in turn \citep{ChapelaCampa2025simod}. The position at both ends sets two tests for tok, whether the simulated log fits the source model under replay and whether a discovered model reads back into the notation, and the demonstration of the study applies both tests with PM4Py as the independent tool.

\section{The tok notation}\label{sec:notation}

\subsection{A document}

\begin{figure}[t]
\centering
\includegraphics[width=0.92\linewidth]{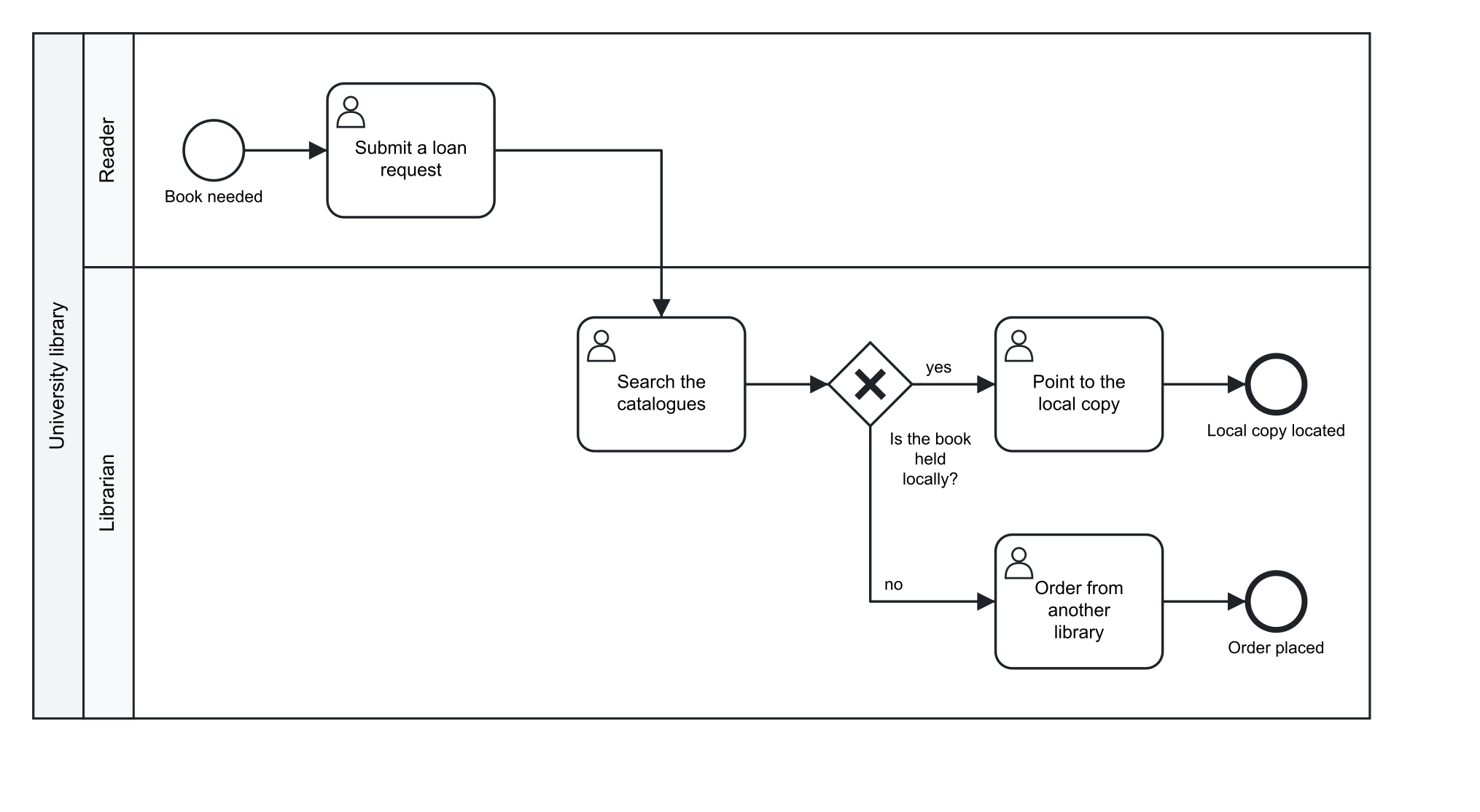}
\caption{The diagram that \texttt{tok svg} draws from the interlibrary-loan document of \cref{lst:loan}, with the automatic layout of the engine, showing one pool, two lanes, an exclusive gateway with two labelled outcomes and two end events.}\label{fig:loan}
\end{figure}

A tok document is a sequence of statements, one statement per line, with blocks marked by indentation. Every statement starts with a keyword, every element has an identifier and an optional quoted label, and attributes follow as \texttt{key=value} pairs or bare flags. The first line declares the language version and the title of the model. \Cref{lst:loan} gives a complete document for an interlibrary loan in one pool with two lanes, and \cref{fig:loan} shows the diagram that the engine draws from the document.

\begin{lstlisting}[caption={A tok document: an interlibrary loan in one pool with two lanes, a decision and two end events.},label={lst:loan}]
tok 1.0 "Interlibrary loan"

pool library "University library"
  lane reader "Reader"
    start need "Book needed"
    user request "Submit a loan request"
  lane librarian "Librarian"
    user catalogues "Search the catalogues"
    xor available "Is the book held locally?"
    user locate "Point to the local copy"
    user order "Order from another library"
    end located "Local copy located"
    end ordered "Order placed"

flow need -> request -> catalogues -> available
flow available -> locate "yes"
flow available -> order "no"
flow locate -> located
flow order -> ordered
\end{lstlisting}

The interlibrary-loan document has 583 bytes and 134 tokens of the o200k tokenizer. The BPMN XML file that \texttt{tok bpmn} writes from the document has 6\,559 bytes and 2\,173 tokens with the diagram, and 2\,330 bytes and 699 tokens without the diagram. The tok text holds the same pools, lanes, nodes, names and flows as the XML file, and the engine computes every coordinate of the diagram.

\subsection{Elements}

The process elements of tok follow the BPMN metamodel one to one \citep{OMG2014BPMN}. Events are written \texttt{start}, \texttt{end}, \texttt{catch}, \texttt{throw} and \texttt{boundary}, with triggers as attributes (\texttt{message}, \texttt{timer}, \texttt{signal}, \texttt{error}, \texttt{escalation}, \texttt{condition}, \texttt{link}, \texttt{compensate}, \texttt{terminate}). Activities are written by their task type (\texttt{task}, \texttt{user}, \texttt{manual}, \texttt{service}, \texttt{script}, \texttt{rule}, \texttt{send}, \texttt{receive}, \texttt{call}) or as sub-processes (\texttt{sub}, \texttt{adhoc}, \texttt{eventsub}, \texttt{transaction}) with an indented block. Gateways are written \texttt{xor}, \texttt{and}, \texttt{or}, \texttt{eventgw} and \texttt{complex}. A \texttt{flow} statement chains sequence flows (\texttt{flow a -> b -> c}) and carries a label, a FEEL condition (\texttt{when=}) or the \texttt{default} flag, while a \texttt{message} statement draws a message flow between pools. Data objects, data stores, item definitions, conversations, choreographies, correlation keys and annotations complete the vocabulary of BPMN 2.0.2.

Decisions are written in the same document as the process, where a \texttt{decision} statement holds a DMN decision table as rows of cells, a boxed context, a list, an invocation of a business knowledge model, or a literal FEEL expression, and \texttt{inputdata}, \texttt{bkm} and \texttt{source} statements form the decision requirements graph of DMN 1.5. A \texttt{rule} task names the decision that the task evaluates. FEEL, the expression language of DMN, is the language of every condition and expression in tok, at conformance level 3 of DMN 1.5 \citep{OMG2024DMN}.

\subsection{Levels, classes and profiles}

The checker accepts a document at one of three teaching levels (\texttt{basic}, \texttt{analytic}, \texttt{full}) and, independently, against one of the conformance classes of BPMN 2.0.2 (descriptive, analytic, common executable, full) \citep{OMG2014BPMN}. A teaching level limits the vocabulary to the elements a course has introduced, and a conformance class limits the vocabulary to the elements a target tool supports. The word \texttt{executable} in the header turns on the profile of Camunda 8 \citep{Camunda8}, under which the checker applies the rules of the engine and the exported BPMN carries the job types, input and output mappings, forms and assignments that the engine needs. Language version 1.0 carries a compatibility promise, according to which every release 1.x reads every 1.0 document with the same meaning.

\subsection{Simulation}

Simulation parameters live in the same document and leave the BPMN export unchanged. A \texttt{simulate} statement sets the number of cases, the seed and the distribution of arrivals, \texttt{resource} statements declare performers with a capacity, and activities take a \texttt{duration} distribution and a performer (\texttt{by=}); flows leaving a gateway take a branch probability (\texttt{p=}) or a FEEL condition over case variables drawn by \texttt{variable} statements. The simulator follows the cases through the diagram with a queue per resource, in the manner of discrete-event business process simulation \citep{Aalst2015simulation}, and writes an event log as CSV and as XES, the IEEE standard format for event logs \citep{IEEE1849}, so that process-mining tools read the log directly. The same document and seed always give the same log.

\section{Classification}\label{sec:classification}

tok is a declarative, domain-specific textual notation \citep{Mernik2005} with a context-free syntax and decidable context conditions, given a translational semantics through the BPMN 2.0.2 and DMN 1.5 metamodels. The classification places the notation in four fields, one field per paragraph, because each field explains a different property of the design, and together the four fields explain why a notation built for people serves language models as well.

In formal language theory, the syntax of tok is line-oriented with blocks marked by indentation, after the offside rule of \citet{Landin1966}. Indentation makes the syntax of tok sensitive to the column of each line, a property beyond the reach of a context-free grammar over characters. A lexer that turns changes of indentation into block-opening and block-closing tokens yields a token language that is context-free in the hierarchy of \citet{Chomsky1956}, and a deterministic top-down parser reads the token language in one pass. Unique identifiers, resolved references and the rules of BPMN 2.0.2 are context conditions over the parse tree, in the sense of attribute grammars \citep{Knuth1968}, so that every condition is decidable and the time the engine needs to check all conditions grows at most linearly with the size of a model. Expressions are delegated to FEEL, an embedded sublanguage with its own grammar. Execution and simulation are separate interpretations of the same text.

In model-driven engineering, tok is an alternative concrete syntax for an existing abstract syntax, the construction which language workbenches such as Xtext and MontiCore support for textual languages \citep{Eysholdt2010xtext,Krahn2010monticore} and which text-based modelling recommends for the sake of versioning, differencing and tooling \citep{Gronniger2014textbased}. The BPMN and DMN metamodels define the abstract syntax, the BPMN diagram and the BPMN XML file are two standard concrete syntaxes of the metamodels, and tok is a third concrete syntax of the same metamodels. The meaning of a tok document is the meaning of the BPMN and DMN models compiled from the document. A notation over an existing metamodel keeps the standard model as the reference and makes an exact translation in both directions possible, in which an imported BPMN file prints as tok and compiles back to a BPMN file with the same semantic content.

In semiotics, the BPMN diagram, the BPMN XML file and the tok text are three expression planes of one content plane, the metamodel, in the terms of \citet{Hjelmslev1961}. The diagram is largely iconic, a picture whose spatial relations stand for the relations of the process, while the tok text is symbolic, a code of conventional signs for the same relations \citep{Peirce1931}. Compiling tok to a diagram is an intersemiotic translation, from one sign system to another of a different kind, and compiling tok to XML is a translation between two codes of the same kind \citep{Jakobson1959}. tok is built as a notational system in the sense of \citet{Goodman1976}, in which the characters of the notation are disjoint and finitely differentiated and every canonical text has one model. The analog part of a drawing, the position of every shape, moves to a separate layout file, so the tok text keeps only the notational part of the model. In the division of \citet{Morris1938}, the grammar of tok is its syntactics, the BPMN meaning its semantics, and the diagnostics, the canonical form and the documents for language-model applications its pragmatics, the design for the users of the notation.

Notation research and the theory of communication supply the criteria for the pragmatic choices. The Physics of Notations judges visual notations by principles such as semiotic clarity and perceptual discriminability \citep{Moody2009}, and the cognitive dimensions of \citet{GreenPetre1996} judge textual and visual notations alike by dimensions which include viscosity, the resistance of a notation to change, hidden dependencies, and progressive evaluation, the possibility to check work at every stage. tok lowers the viscosity of a model, since an edit of one line changes one line, makes the dependencies of a flow visible in the flow statement, and gives progressive evaluation through a checker which answers every text in about 20 milliseconds.

In the account of grounding by \citet{ClarkBrennan1991}, two parties reach a shared understanding by exchanging evidence of understanding at the least collaborative effort, and a party which receives evidence of a misunderstanding repairs the utterance at the point which the evidence names. The checker of tok supplies such evidence for every text in the form of a diagnostic with a code, a line and an element, to a person in an editor and to a language model in a loop alike, and the line-oriented syntax keeps the evidence local, since a statement occupies one line and a reference names one identifier. The repair of a fault therefore touches one line, a property which research on error messages for programmers ties to the usefulness of a message \citep{Marceau2011measuring,Becker2019compiler}. In the terms of \citet{HutchinsHollanNorman1985}, a localised diagnostic narrows the gulf of evaluation, the distance between the state of the system and the understanding of the user, which for a person means a visible fault in the editor and for a language model means a short message to answer. The reference condition of the study measures how much grounding one page of reference and the messages of the checker give a model which has never seen the notation, and the agent condition measures what a skill document and tools called by the model add. A process model in tok is a boundary object in the sense of \citet{StarGriesemer1989}, a text with enough structure to be read by analysts, students, process engines through the exported BPMN, and language models, and with enough plasticity to be written by each of them.

\section{The engine}\label{sec:engine}

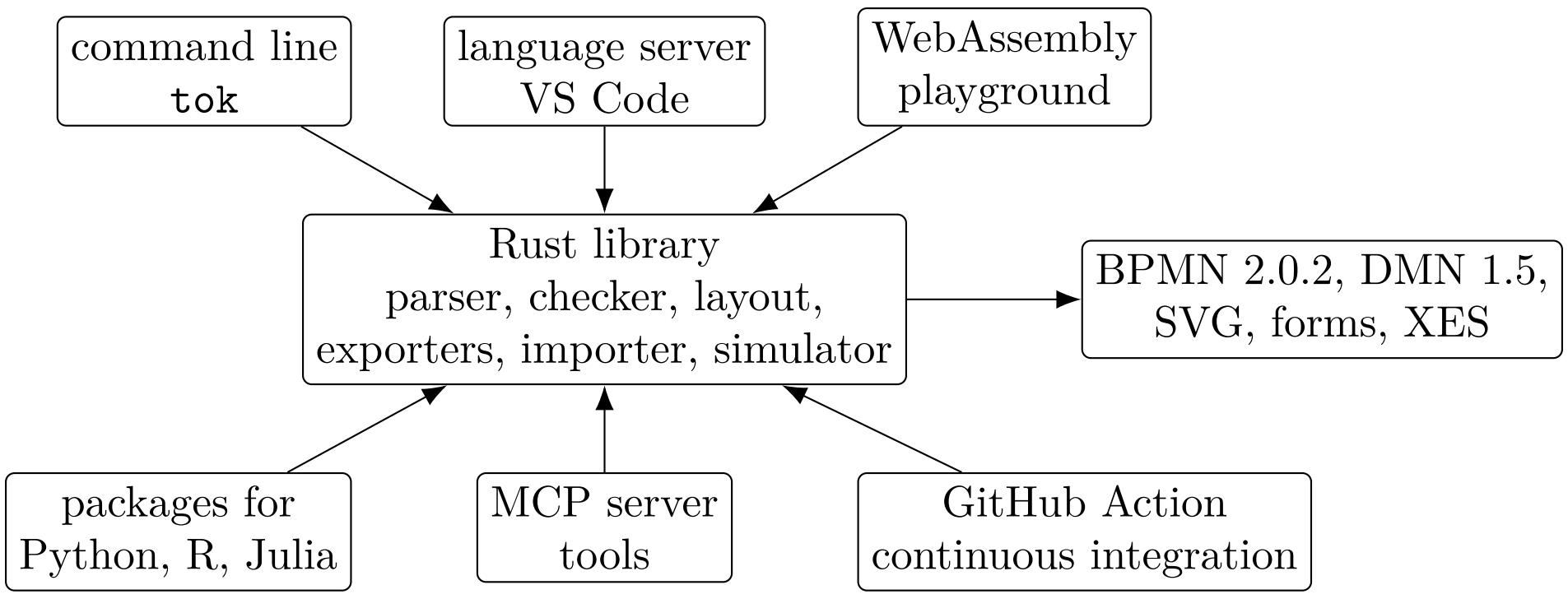
\begin{figure}[t]
\centering
\begin{tikzpicture}[node distance=4mm and 6mm, box/.style={draw,rounded corners=2pt,align=center,font=\small,minimum height=8mm,inner sep=3pt}, arr/.style={-{Latex[length=2mm]}}]
\node[box,minimum width=36mm] (core) {Rust library\\parser, checker, layout,\\exporters, importer, simulator};
\node[box,above left=7mm and -4mm of core] (cli) {command line\\\texttt{tok}};
\node[box,above=7mm of core] (lsp) {language server\\VS Code};
\node[box,above right=7mm and -4mm of core] (wasm) {WebAssembly\\playground};
\node[box,below left=7mm and -4mm of core] (pkg) {packages for\\Python, R, Julia};
\node[box,below=7mm of core] (mcp) {MCP server\\tools};
\node[box,below right=7mm and -4mm of core] (gha) {GitHub Action\\continuous integration};
\foreach \n in {cli,lsp,wasm,pkg,mcp,gha} \draw[arr] (\n) -- (core);
\node[box,right=14mm of core,minimum width=30mm] (std) {BPMN 2.0.2, DMN 1.5,\\SVG, forms, XES};
\draw[arr] (core) -- (std);
\end{tikzpicture}
\caption{Architecture of tok, in which every interface calls one Rust library and the library writes the standard files.}\label{fig:architecture}
\end{figure}

The engine is written once, in Rust \citep{Matsakis2014rust}, and every interface calls the same library (\cref{fig:architecture}). The command-line program \texttt{tok} offers the operations \texttt{check}, \texttt{fmt}, \texttt{bpmn}, \texttt{svg}, \texttt{export}, \texttt{import}, \texttt{sim} and \texttt{eval}. The language server gives an editor the diagnostics, the canonical form, completion, hover information and an outline through the Language Server Protocol \citep{Microsoft2022lsp}, and a VS Code extension uses the language server. The WebAssembly build \citep{Haas2017webassembly} runs the engine in the browser playground. The packages for Python and R bind the library as native extensions, the Julia package calls the library through its C interface, and the MCP server offers the operations as tools to language-model applications. One implementation gives one behaviour, so that a diagnostic has the same code, line and message in the editor, in R, in the browser and through MCP.

The checker reports layered diagnostics, whose layers are tokens, statements, references, the rules of BPMN 2.0.2 with their clauses \citep{OMG2014BPMN}, the profile of Camunda 8 \citep{Camunda8}, decisions, FEEL expressions \citep{OMG2024DMN} and simulation parameters. Every diagnostic carries a code, a severity (error, warning or note), the line, the element and a message in English or Polish. Errors stop the export, and warnings mark a model that compiles to valid XML but breaks a rule of the standard. The interlibrary-loan document without the flow from \texttt{locate} to \texttt{located} gives the two warnings of the following check.

\begin{lstlisting}
$ tok check interlibrary-loan-error.tok
interlibrary-loan-error.tok:10: warning W042: path without an end event (element without outgoing flow) (locate)
interlibrary-loan-error.tok:12: warning W041: element not reachable from a start event (located)
interlibrary-loan-error.tok: 0 error(s), 2 warning(s), 0 note(s)
\end{lstlisting}

The warnings of the BPMN rules cover every correctness rule of bpmnlint, the linter of the bpmn.io tools \citep{bpmnlint}, so a model with no errors and no warnings of \texttt{tok check} also passes the linter. The same rules check BPMN XML files from any tool, and \texttt{tok check FILE.bpmn} reports the findings with the line of each element.

The importer reads BPMN and DMN files and prints the files as tok. The geometry of an imported drawing goes to a layout file beside the text, and the exporters reuse the drawing when the layout file is present, so a model drawn in a graphical tool keeps its drawing through the round trip while the tok text holds only the model. The canonical form, printed by \texttt{tok fmt}, gives one text per model and clean differences under version control.

The MCP server \citep{MCP2025} exposes the operations \texttt{check}, \texttt{check\_bpmn}, \texttt{fmt}, \texttt{export}, \texttt{import}, \texttt{sim}, \texttt{eval} and \texttt{reference} as tools, and two skill documents describe the modelling procedure for language-model applications, one for tok and one for BPMN XML. The tok skill sets the order of work, which leads from the process description through a model, the check and the correction of every error and warning to the canonical form and the export.

A pure-Python implementation of the language core serves as an executable specification. Parity tests compare the canonical text, the semantic BPMN XML and the diagnostic codes of the Rust engine and of the Python implementation byte for byte on a conformance corpus of 67 documents.

\section{Study design}\label{sec:design}

\subsection{Questions and software}

The study answers six numbered questions about tok. Question 1 asks how many fewer tokens a model takes in tok than in BPMN XML or DMN XML, and what the conversion of tool files to tok keeps. The second question, Question 2, is whether the tok checker finds injected modelling faults that the XML validators find and points to the faulty element, while Question 3 asks whether language models write valid process models in tok as often as in BPMN XML, with how many tokens and how faithful to the text, from a one-page reference and as agents with tools. Process mining is the subject of Question 4, with tok as the generator of an event log from a model and as the reader of a discovered model, and Questions 5 and 6 ask how fast the engine is and whether the released engine passes its conformance suites.

The study uses two frozen releases of tok, both published with archived binaries \citep{Deja2026tok}. Release 1.0.2 serves the size measurement, the fault injection and the reference condition of the language-model study. Release 1.0.3 serves the agent condition of the language-model study, the process-mining demonstration, the timing and the conformance suites, and carries the MCP tools and the skill document for language-model applications and four further BPMN warnings (an implicit split, a sub-process without a start or an end event, and an event with several definitions). The comparison tools are pinned at bpmnlint 12.0.0 with bpmn-moddle 10.3.1 \citep{bpmnlint}, the OMG XML Schema of BPMN 2.0, and PM4Py 2.7.15 \citep{Berti2023PM4Py}.

\subsection{Corpora}

\Cref{tab:corpora} lists the corpora, of which corpus N holds the tok models of the project, namely the examples of the repository, the conformance cases, and the examples of the guide and of a course, deduplicated and without errors. Corpus V holds the BPMN files that 37 modelling tools exported for the test suite of the BPMN Model Interchange Working Group \citep{MIWG}, and corpus R the 22 reference models of the same suite. Corpus C holds 16 BPMN files drawn in Camunda Modeler for a university course, and corpus D the DMN models of the DMN Technology Compatibility Kit \citep{DMNTCK}. The texts of the language-model study come from two datasets. PET holds 45 process descriptions with expert annotations of activities, actors, gateways and flow relations \citep{Bellan2022EDOC,Bellan2023PET}. The PMo dataset holds 55 process descriptions, each with a BPMN model, from five sources \citep{Brissard2025PMo}, and the seven PMo descriptions that repeat PET texts are left out.

\begin{table}[t]
\centering
\small
\caption{Corpora of the study, with the number of items and the questions each corpus serves.}\label{tab:corpora}
\begin{tabularx}{\linewidth}{@{}l X r l@{}}
\toprule
Corpus & Contents & Items & Questions\\
\midrule
N & tok models of the project, without errors & 93 & 1, 2, 5\\
V & BPMN files exported by 37 tools (MIWG test suite) & 893 & 1\\
R & MIWG reference models & 22 & 1, 5, 6\\
C & BPMN files drawn in Camunda Modeler & 16 & 1, 6\\
D & DMN models of the DMN TCK, levels 2 and 3 & 143 & 1\\
PET & process descriptions with expert annotations (PET 1.1) & 42 & 3\\
PMo & process descriptions with BPMN models (PMo 1.0.0) & 45 & 3\\
PMo models & BPMN models of the PMo descriptions & 48 & 4\\
\bottomrule
\end{tabularx}
\end{table}

\subsection{Size and fidelity}

The size measurement counts characters, bytes and the tokens of five open subword tokenizers of the byte-pair kind \citep{Sennrich2016}, namely o200k and cl100k of tiktoken \citep{tiktoken}, Qwen3, DeepSeek-V3 and Mistral NeMo. For corpus N the measurement compares the tok text with the standard files that the engine exports, with and without diagrams, as written and minified. For corpora V, R and C the measurement compares the file as exported by the tool, the file without the diagram, and the smallest standard BPMN of the same model (without the diagram, without vendor extensions, minified) with the tok text that \texttt{tok import} writes. The ratio of XML tokens to tok tokens is computed per file and summarised by the geometric mean with a 95\,\% bootstrap interval (10\,000 resamples of files, seed 20261029), the median and the quartiles.

The fidelity measurement reads the original file and the BPMN file that tok writes back after the import with an independent reader built on lxml. The reader compares the kinds of flow nodes, the names, the sequence flows, the pools, the lanes and the message flows. A file whose import has checker errors only because the tool marked the process as executable is exported again with the executable mark removed, and both routes are reported.

\subsection{Fault injection}

The fault injection starts from the 85 models of corpus N with a start event, an end event and at least four flow nodes. Thirteen operators inject one fault each, namely a flow to an undefined node, a duplicate identifier, a missing start event, a missing end event, a disconnected activity, an exclusive split without conditions or a default, an exclusive split joined by a parallel join, a boundary event on an undefined activity, a start event with an incoming flow, an end event with an outgoing flow, a sequence flow between two pools, a malformed condition and a flow from an activity to itself. Every operator edits the tok text and, as the same edit, the semantic BPMN XML of the unmodified model, at the first, middle and last applicable element, which gives 1\,411 faulty variants in all. The validators are \texttt{tok check} at the full level, the OMG XML Schema \citep{OMG2014BPMN}, bpmnlint with its correctness rules and bpmnlint with its recommended rules \citep{bpmnlint}. A fault is detected when a validator reports a new error, signalled when the validator reports a new error or a new warning, and localised when a new diagnostic names the edited element or points to its line. Intervals of the overall rates come from resampling base models.

\subsection{Language models}

Four language models that work as agents with tools write the models, which are Claude Haiku 5.5 and Claude Sonnet 5.5 through the Claude command line, and the open-weight models gpt-oss 120B and Beam through the Ollama service. Each model writes a process model for each text in six arms, which cross the two representations with three conditions. In the condition \emph{reference} the model gets a one-page reference of the representation with one worked example, the answer is checked, the messages are sent back, and the model has at most four attempts; the check of BPMN XML is the OMG XML Schema \citep{OMG2014BPMN}, the correctness rules of bpmnlint \citep{bpmnlint} and a minimum of structure, and the check of tok is \texttt{tok check} of release 1.0.2, whose errors and warnings are sent back. In the two agent conditions the model gets a skill document as its system prompt, calls tools of the MCP server of tok 1.0.3 as often as the model decides, and has at most 12 model calls. In the condition \emph{agent, distributed skill} the skill is the skill document distributed with tok 1.0.3 for the representation: for tok, a document of about 1\,950 words which names the tools \texttt{check}, \texttt{fmt} and \texttt{reference}, through which the model reads the one-page reference, the specification or the list of diagnostics; for BPMN XML, a document of about 1\,630 words which names the tool \texttt{check\_bpmn}, lets the model justify a warning in place of fixing it, and gets the one-page reference of BPMN XML in place of the reference files of the skill. In the condition \emph{agent, lean skill} the skill is a document of five steps and about 180 words, the same for both representations up to the names of the elements and of the tool, which carries the one-page reference of the representation, names one checker tool (\texttt{check} for tok, \texttt{check\_bpmn} for BPMN XML) and asks the model to fix every error and every warning before answering. Every arm uses the same instruction, which asks for one pool, one lane per actor, tasks named with the words of the text, gateways for decisions and parallel work, and a start and an end event on every path. The XML arms ask for the semantic model without the diagram, the cheapest correct XML. The language models know BPMN XML from their training and meet tok for the first time in the prompt.

The 42 PET texts run twice per model in the reference condition and once in the agent conditions, while the 45 PMo texts run once per model in every arm. Runs follow an order shuffled with seed 20261030. A harness test on three further texts of each dataset, before the study, confirmed that the harness parses the answers and runs the validators. A probe of the tool-call channel of Beam, whose calls of \texttt{check\_bpmn} arrive without the document in the harness test, asks the model to pass a given document unchanged to a checker tool and records whether the document arrives, for one tok model and its BPMN XML export, for the XML cut to the length of the tok text, for the tok text repeated to the length of the XML, for a plain text of the XML length without quote characters, and for the XML through four variants of the definition of \texttt{check\_bpmn} and through \texttt{check}, three repeats each.

One judge scores every final model on BPMN XML, which for tok is the BPMN file that \texttt{tok bpmn} writes from the answer and for XML the answer itself. A model is valid when the file is well-formed and valid against the OMG schema, has no error of the bpmnlint correctness rules, has a start and an end event, and has only flows between known nodes. The judge is independent of the tok checker. The cost of a run is the number of input and output tokens reported by the provider for all calls of the run, including cache reads and cache writes for the Claude models, and the wall time. The fidelity of a valid model compares the model with the gold standard of the text, in that activity names are matched one to one by the Hungarian method \citep{Kuhn1955hungarian} on a token-set similarity of at least 0.5 computed with RapidFuzz \citep{rapidfuzz}, and flow F1 compares the ordering of activities through gateways and events in the model with the gold ordering. The gold ordering of a PET text comes from the expert annotation, and the gold ordering of a PMo text from the BPMN model of the dataset. The three primary outcomes are validity, total tokens per run and flow F1 with an invalid model scored 0.

The analysis compares the arms within each model and dataset, with the text as the unit. For each text the outcome is averaged over the repetitions of an arm. The four primary comparisons are \emph{tok, reference} against \emph{BPMN XML, reference} (the representation under the reference condition), \emph{tok, agent, lean skill} against \emph{BPMN XML, agent, lean skill} (the representation under the same skill), \emph{tok, agent, lean skill} against \emph{tok, agent, distributed skill} (the skill design within tok), and \emph{tok, agent, lean skill} against \emph{tok, reference} (the agent loop against the reference loop within tok). Each comparison reports the mean difference with a 95\,\% bootstrap interval (10\,000 resamples of texts, seed 20261029) and the Wilcoxon signed-rank test \citep{Wilcoxon1945,Efron1979}. Holm's procedure adjusts the $p$ values of the 96 primary tests together (four models, two datasets, four comparisons, three outcomes) \citep{Holm1979}. The remaining pairs of arms, among them the two distributed skills against each other, are reported with the same intervals and without a test.

\subsection{Process mining}

The process-mining demonstration takes the 48 BPMN models of the PMo dataset. For each model, \texttt{tok import} reads the BPMN file and a fixed rule adds the simulation (every activity takes an exponential duration with a mean of 10 minutes, every flow leaving an exclusive gateway without a condition takes an equal share of probability, every flow leaving an inclusive gateway a probability of 0.5, and 500 cases arrive with exponential gaps of one hour on average, seed 20261031), after which \texttt{tok sim} writes the event log as XES. PM4Py replays the log on the original BPMN file converted to a Petri net, by token-based replay \citep{Berti2021tbr}, and reports fitness and precision. The inductive miner of PM4Py \citep{Leemans2013}, with a noise threshold of 0, discovers a model from the log and writes the model as BPMN. \texttt{tok import} reads the discovered file, \texttt{tok check} checks the imported model, and \texttt{tok bpmn} writes the model again for a comparison of the round trip. The ordering of activities of the discovered model is compared with the ordering of the original model by precision, recall and F1. PM4Py also reads the BPMN file that \texttt{tok bpmn} writes from each simulation model, and the demonstration records how many sequence flows PM4Py finds.

\subsection{Performance and conformance}

The timing runs the released program on every model of corpus N (check, canonical form, BPMN with layout, SVG), on the import of corpus R, and on generated models of 10 to 3\,000 flow nodes, each operation a separate start of the program, five runs, median, on a Windows laptop. The conformance measurement runs the suites of the tagged sources of release 1.0.3, namely the DMN Technology Compatibility Kit at compliance levels 2 and 3, the import, printing and export of the 22 MIWG reference models and the 16 Camunda Modeler files, and the 67 cases of the tok conformance corpus.

\section{Results}\label{sec:results}

\subsection{Size and fidelity}

A model in tok takes two to four times fewer tokens than the smallest standard XML of the same model, and six to twelve times fewer than XML with a diagram (\cref{tab:size}). For the 893 files that 37 tools exported for the MIWG test suite, the geometric mean ratio is 2.18 (95\,\% interval 2.15--2.21) against the smallest standard BPMN and 7.71 (7.48--7.96) against the files as exported. For the 93 tok models of the project, the standard files that the engine exports take 3.90 times the tokens of the tok text without diagrams and 11.42 times with diagrams. DMN decisions take 2.49 times the tokens of tok in the smallest DMN XML and 3.83 times in the files of the DMN TCK. The ratios of the other four tokenizers lie between 11\,\% below and 23\,\% above the o200k ratios, and the ratio of cl100k within 3\,\%.

\begin{table}[t]
\centering
\small
\caption{Tokens of XML divided by tokens of tok (o200k tokenizer), as the geometric mean over files with a 95\,\% bootstrap interval. The smallest XML has no diagram and no vendor extensions and is minified. Corpus N compares the tok text with the files the engine exports, and corpora V, R, C and D compare the tool file with the tok text of its import.}\label{tab:size}
\begin{tabular}{@{}l r l r@{}}
\toprule
Corpus & Files & XML variant & Ratio [95\,\% interval]\\
\midrule
N, tok models & 93 & smallest standard files & 3.90 [3.73, 4.08]\\
 & & standard files with diagrams & 11.42 [10.79, 12.06]\\
V, files of 37 tools & 893 & smallest BPMN & 2.18 [2.15, 2.21]\\
 & & file as exported & 7.71 [7.48, 7.96]\\
R, MIWG reference models & 22 & smallest BPMN & 2.14 [1.98, 2.33]\\
 & & file as exported & 7.05 [6.04, 8.39]\\
C, Camunda Modeler files & 16 & smallest BPMN & 2.24 [2.17, 2.31]\\
 & & file as exported & 6.13 [5.92, 6.34]\\
D, DMN TCK models & 143 & smallest DMN & 2.49 [2.32, 2.68]\\
 & & file as published & 3.83 [3.49, 4.22]\\
\bottomrule
\end{tabular}
\end{table}

The conversion of tool files to tok keeps the content of every file that the checker accepts. Of the 931 BPMN files of corpora V, R and C, 681 files import and export without checker errors, 75 of the 681 files after the executable mark of the tool is removed. The independent reader finds every flow-node kind, every name, every sequence flow, every pool and every lane of the 681 originals in the exported files, and 99.6\,\% of the message flows of corpus V. The remaining 250 files carry checker errors, most often a condition that FEEL cannot parse (code F002, 121 files) and a node outside every lane of a pool with lanes (R030, 66 files).

\subsection{Fault detection}

The tok checker signals 98.2\,\% of the 1\,411 injected faults with an error or a warning (95\,\% interval 97.4--99.0), and the XML Schema together with the correctness rules of bpmnlint signals 85.8\,\% (83.9--87.7) (\cref{tab:faults}). The tok checker reports 56.5\,\% of the faults as errors, which stop the export, and the remaining signalled faults as warnings. Of the faults that the tok checker reports, 99.7\,\% carry a diagnostic that names the faulty element or its line, against 87.1\,\% for the XML validators. The XML Schema alone detects 22.7\,\% of the faults, mostly duplicate identifiers.

\begin{table}[t]
\centering
\small
\caption{Detection of 1\,411 injected faults in 85 base models, given as the share reported as an error (detected), as an error or a warning (signalled), and share of reported faults whose diagnostic names the faulty element or line (localised). The 95\,\% intervals come from resampling base models.}\label{tab:faults}
\begin{tabular}{@{}l r r r@{}}
\toprule
Validator & Detected & Signalled & Localised\\
\midrule
\texttt{tok check} & 56.5\,\% [54.6, 58.3] & 98.2\,\% [97.4, 99.0] & 99.7\,\%\\
XML Schema & 22.7\,\% [20.4, 25.3] & 22.7\,\% [20.4, 25.3] & 100\,\%\\
bpmnlint, correctness rules & 65.8\,\% [63.5, 68.1] & 84.0\,\% [82.2, 85.7] & 83.9\,\%\\
bpmnlint, recommended rules & 65.8\,\% [63.5, 68.1] & 94.1\,\% [92.2, 95.9] & 83.9\,\%\\
XML Schema or bpmnlint correctness & 82.6\,\% [80.3, 85.0] & 85.8\,\% [83.9, 87.7] & 87.1\,\%\\
\bottomrule
\end{tabular}
\end{table}

Two operators mark the limits of the tok checker. An exclusive split without conditions or a default is signalled in 15 of 27 variants (55.6\,\%), and an exclusive split joined by a parallel join, a deadlock, in 1 of 5 variants. The other eleven operators are signalled in 96.5 to 100\,\% of their variants.

\subsection{Language models}

\begin{figure}[t]
\centering
\includegraphics[width=\linewidth]{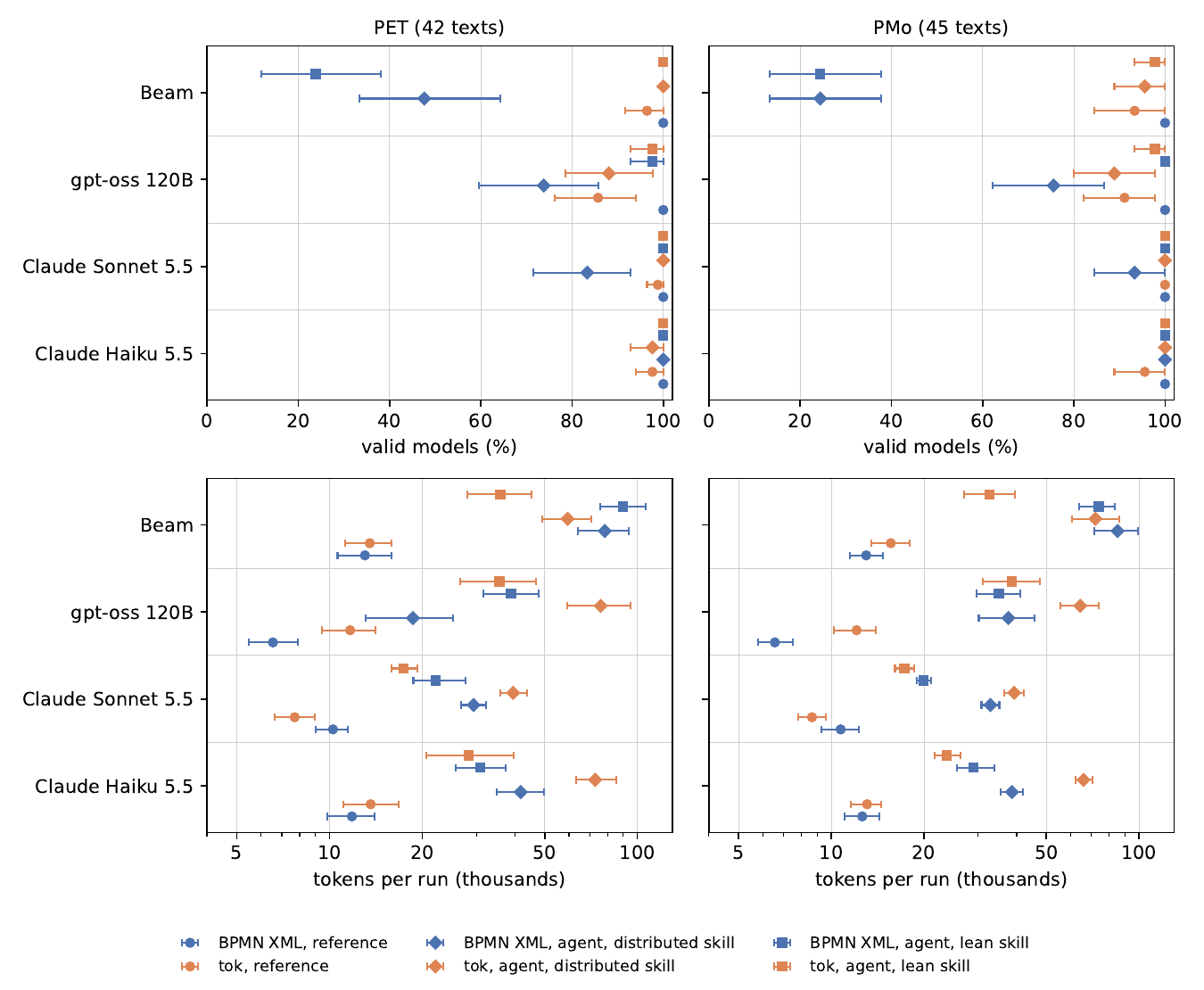}
\caption{Share of valid models (top) and total tokens per run in thousands (bottom, logarithmic axis) for each model and arm, on the 42 PET texts (left) and the 45 PMo texts (right). Blue marks BPMN XML and orange marks tok; a circle marks the reference condition, a diamond the agent with the distributed skill and a square the agent with the lean skill. The points mark means over texts and the bars 95\,\% bootstrap intervals over texts.}\label{fig:models}
\end{figure}

The four language models write valid models in every arm except the two BPMN XML agent arms of Beam, and the arms differ in reliability for the weaker models and in tokens for every model (\cref{tab:models,fig:models}). In BPMN XML from the one-page reference, every model writes a valid model for every text of both sets. In tok from the one-page reference, the share of valid models is 97.6 and 95.6\,\% for Claude Haiku 5.5 on PET and PMo, 98.8 and 100\,\% for Claude Sonnet 5.5, 96.4 and 93.3\,\% for Beam, and 85.7 and 91.1\,\% for gpt-oss 120B. As agents with the lean skill and one checker tool, Claude Sonnet 5.5 and Claude Haiku 5.5 write a valid model for every text of both sets in both representations, gpt-oss 120B for 97.6 and 97.8\,\% of the texts in tok and for 97.6 and 100\,\% in BPMN XML, and Beam for 100 and 97.8\,\% in tok. With the distributed skills, the agents in tok reach 95.6 to 100\,\% for Claude Haiku 5.5, Claude Sonnet 5.5 and Beam and 88.1 and 88.9\,\% for gpt-oss 120B, and the agents in BPMN XML reach 100\,\% for Claude Haiku 5.5, 83.3 and 93.3\,\% for Claude Sonnet 5.5 and 73.8 and 75.6\,\% for gpt-oss 120B, since the distributed BPMN skill lets the model justify a warning in place of a fix, and the final models of Claude Sonnet 5.5 keep a branch without a condition next to a default flow in 7 and 3 runs, which bpmnlint counts as an error. After Holm's adjustment over the 96 primary tests, the only significant differences in validity are the two comparisons of Beam in BPMN XML reported below, and the largest other difference among the primary tests, gpt-oss 120B in tok against BPMN XML from the reference on PET, is $-0.143$ (95\,\% interval $-0.238$ to $-0.060$, $p_\text{Holm}=0.64$).

\begin{table}[tp]
\centering
\footnotesize
\caption{Language models writing process models from the 42 PET texts and the 45 PMo texts in six arms, which cross the representation (BPMN XML, tok) with the condition. The reference condition gives a one-page reference and the checker's messages, with at most four attempts. The agent with the distributed skill gets the skill document distributed with tok 1.0.3 for the representation and its tools, and the agent with the lean skill gets a five-step skill with the one-page reference and one checker tool, both with at most 12 model calls. Share of valid models, mean total and output tokens per run in thousands (as reported by the provider, all calls of a run), and mean flow F1 against the gold ordering with invalid models scored 0. Means over texts, with the repetitions of a text averaged.}\label{tab:models}
\begin{tabular}{@{}l l l r r r r@{}}
\toprule
Set & Model & Arm & Valid (\%) & Tokens & Output & Flow F1\\
\midrule
PET & Claude Haiku 5.5 & BPMN XML, reference & 100.0 & 11.8 & 6.2 & 0.530\\
 &  & tok, reference & 97.6 & 13.6 & 5.3 & 0.510\\
 &  & BPMN XML, agent, distributed skill & 100.0 & 41.8 & 10.8 & 0.516\\
 &  & tok, agent, distributed skill & 97.6 & 72.9 & 7.2 & 0.483\\
 &  & BPMN XML, agent, lean skill & 100.0 & 30.9 & 8.8 & 0.504\\
 &  & tok, agent, lean skill & 100.0 & 28.3 & 4.7 & 0.547\\
\addlinespace
 & Claude Sonnet 5.5 & BPMN XML, reference & 100.0 & 10.3 & 3.6 & 0.493\\
 &  & tok, reference & 98.8 & 7.7 & 1.1 & 0.503\\
 &  & BPMN XML, agent, distributed skill & 83.3 & 29.4 & 5.7 & 0.448\\
 &  & tok, agent, distributed skill & 100.0 & 39.5 & 1.8 & 0.496\\
 &  & BPMN XML, agent, lean skill & 100.0 & 22.2 & 5.0 & 0.488\\
 &  & tok, agent, lean skill & 100.0 & 17.4 & 1.3 & 0.512\\
\addlinespace
 & gpt-oss 120B & BPMN XML, reference & 100.0 & 6.5 & 3.4 & 0.462\\
 &  & tok, reference & 85.7 & 11.7 & 3.8 & 0.422\\
 &  & BPMN XML, agent, distributed skill & 73.8 & 18.7 & 4.3 & 0.422\\
 &  & tok, agent, distributed skill & 88.1 & 76.1 & 5.5 & 0.432\\
 &  & BPMN XML, agent, lean skill & 97.6 & 38.9 & 7.8 & 0.431\\
 &  & tok, agent, lean skill & 97.6 & 35.7 & 4.2 & 0.436\\
\addlinespace
 & Beam & BPMN XML, reference & 100.0 & 13.0 & 9.3 & 0.543\\
 &  & tok, reference & 96.4 & 13.5 & 7.9 & 0.530\\
 &  & BPMN XML, agent, distributed skill & 47.6 & 78.4 & 11.4 & 0.261\\
 &  & tok, agent, distributed skill & 100.0 & 59.4 & 6.1 & 0.466\\
 &  & BPMN XML, agent, lean skill & 23.8 & 89.8 & 12.5 & 0.130\\
 &  & tok, agent, lean skill & 100.0 & 35.9 & 6.6 & 0.535\\
\midrule
PMo & Claude Haiku 5.5 & BPMN XML, reference & 100.0 & 12.6 & 6.5 & 0.371\\
 &  & tok, reference & 95.6 & 13.1 & 3.8 & 0.355\\
 &  & BPMN XML, agent, distributed skill & 100.0 & 38.6 & 10.3 & 0.379\\
 &  & tok, agent, distributed skill & 100.0 & 66.0 & 6.8 & 0.398\\
 &  & BPMN XML, agent, lean skill & 100.0 & 29.0 & 8.5 & 0.388\\
 &  & tok, agent, lean skill & 100.0 & 23.7 & 3.9 & 0.388\\
\addlinespace
 & Claude Sonnet 5.5 & BPMN XML, reference & 100.0 & 10.7 & 3.8 & 0.425\\
 &  & tok, reference & 100.0 & 8.7 & 1.2 & 0.425\\
 &  & BPMN XML, agent, distributed skill & 93.3 & 32.9 & 6.5 & 0.418\\
 &  & tok, agent, distributed skill & 100.0 & 39.3 & 1.7 & 0.421\\
 &  & BPMN XML, agent, lean skill & 100.0 & 20.0 & 4.8 & 0.404\\
 &  & tok, agent, lean skill & 100.0 & 17.3 & 1.2 & 0.380\\
\addlinespace
 & gpt-oss 120B & BPMN XML, reference & 100.0 & 6.6 & 3.4 & 0.347\\
 &  & tok, reference & 91.1 & 12.1 & 3.8 & 0.349\\
 &  & BPMN XML, agent, distributed skill & 75.6 & 37.6 & 7.2 & 0.260\\
 &  & tok, agent, distributed skill & 88.9 & 64.4 & 5.1 & 0.329\\
 &  & BPMN XML, agent, lean skill & 100.0 & 35.0 & 7.6 & 0.380\\
 &  & tok, agent, lean skill & 97.8 & 38.6 & 4.3 & 0.379\\
\addlinespace
 & Beam & BPMN XML, reference & 100.0 & 13.0 & 8.8 & 0.415\\
 &  & tok, reference & 93.3 & 15.6 & 8.6 & 0.380\\
 &  & BPMN XML, agent, distributed skill & 24.4 & 85.1 & 11.2 & 0.108\\
 &  & tok, agent, distributed skill & 95.6 & 72.1 & 6.5 & 0.410\\
 &  & BPMN XML, agent, lean skill & 24.4 & 74.0 & 11.8 & 0.075\\
 &  & tok, agent, lean skill & 97.8 & 32.6 & 6.2 & 0.398\\
\bottomrule
\end{tabular}
\end{table}

Claude Sonnet 5.5 writes tok with fewer tokens than BPMN XML from the reference. On PET a run in tok takes 7\,709 tokens against 10\,263 in BPMN XML, a difference of $-2\,553$ tokens (95\,\% interval $-3\,525$ to $-1\,555$, $p_\text{Holm}<0.001$). On PMo the run takes 8\,659 tokens against 10\,730, a difference of $-2\,071$ ($-3\,733$ to $-518$, unadjusted $p=0.012$, $p_\text{Holm}=0.77$). The output of Claude Sonnet 5.5 shrinks most, from 3\,579 to 1\,142 tokens on PET and from 3\,804 to 1\,206 on PMo. For Claude Haiku 5.5 and Beam the total tokens of the two representations lie within 3\,000 of each other, and gpt-oss 120B uses more tokens for tok than for BPMN XML (11\,670 against 6\,550 on PET, 12\,089 against 6\,559 on PMo, $p_\text{Holm}<0.001$ for both sets), because the model needs more attempts for tok.

The design of the skill decides the cost of the agent loop. With the distributed skill of tok, an agent run takes 39.5 and 39.3 thousand tokens for Claude Sonnet 5.5 on PET and PMo, 72.9 and 66.0 thousand for Claude Haiku 5.5, 76.1 and 64.4 thousand for gpt-oss 120B and 59.4 and 72.1 thousand for Beam, because the skill sends the model to the \texttt{reference} tool, whose result returns with every later turn of the conversation, and the agents read the reference about once per run. With the lean skill the same agents take 17.4 and 17.3 thousand tokens (Claude Sonnet 5.5), 28.3 and 23.7 thousand (Claude Haiku 5.5), 35.7 and 38.6 thousand (gpt-oss 120B) and 35.9 and 32.6 thousand (Beam), a reduction of 22.0 to 44.6 thousand tokens per run, or 40 to 64\,\%, which is significant for every model on both sets ($p_\text{Holm}\le0.008$). The validity stays equal for Claude Sonnet 5.5 and stays equal or rises for the other three models, most for gpt-oss 120B, from 88.1 and 88.9 to 97.6 and 97.8\,\% ($p_\text{Holm}=1$).

Under the lean skill, tok costs the Claude models fewer tokens than BPMN XML. An agent run of Claude Sonnet 5.5 takes 17.4 thousand tokens in tok against 22.2 thousand in BPMN XML on PET, a difference of $-4\,720$ tokens (95\,\% interval $-8\,485$ to $-2\,133$, $p_\text{Holm}=0.006$), and 17.3 against 20.0 thousand on PMo, a difference of $-2\,700$ ($-3\,749$ to $-1\,569$, $p_\text{Holm}=0.017$). A run of Claude Haiku 5.5 takes 28.3 against 30.9 thousand tokens on PET, a difference of $-2\,642$ whose interval includes zero ($-7\,667$ to $3\,426$, $p_\text{Holm}=0.074$), and 23.7 against 29.0 thousand on PMo, a difference of $-5\,307$ ($-8\,242$ to $-2\,795$, $p_\text{Holm}=0.008$). For gpt-oss 120B the two representations cost the same under the lean skill, with 35.7 against 38.9 thousand tokens on PET and 38.6 against 35.0 thousand on PMo ($p_\text{Holm}=1$). The output of the agents, the part of the cost which the representation controls directly, is 26\,\% of the BPMN XML output for Claude Sonnet 5.5 (1.3 against 5.0 thousand tokens on PET, 1.2 against 4.8 thousand on PMo), 54 and 46\,\% for Claude Haiku 5.5, and 54 and 57\,\% for gpt-oss 120B. A run of Claude Sonnet 5.5 as an agent with the lean skill costs \$0.024 in tok against \$0.069 in BPMN XML on PET and \$0.022 against \$0.065 on PMo, and a run of Claude Haiku 5.5 costs \$0.004 against \$0.006 and \$0.003 against \$0.006.

Under the distributed skills the order of the representations in tokens reverses. The tok agent takes 10.1 and 6.4 thousand tokens more than the BPMN XML agent for Claude Sonnet 5.5, 31.1 and 27.4 thousand more for Claude Haiku 5.5 and 57.4 and 26.8 thousand more for gpt-oss 120B, which is the price of the reference which the tok skill makes the model read, while the validity of the tok agent stays equal or higher (Claude Sonnet 5.5 100 against 83.3 and 93.3\,\%, gpt-oss 120B 88.1 and 88.9 against 73.8 and 75.6\,\%). The lean BPMN skill, which asks the model to fix every warning, brings the BPMN XML agents to 100\,\% for Claude Sonnet 5.5 and to 97.6 and 100\,\% for gpt-oss 120B, and cuts the tokens of the Claude models by 7.2 and 12.9 thousand for Claude Sonnet 5.5 and by 10.9 and 9.7 thousand for Claude Haiku 5.5. Under the distributed BPMN skill gpt-oss 120B calls the checker in 0.8 and 2.1 calls per run and leaves warnings and errors in its final models, whereas under the lean skill the same model calls the checker in 3.0 and 2.8 calls per run, so that its tokens rise on PET from 18.7 to 38.9 thousand and stay on PMo at 37.6 and 35.0 thousand.

The agent loop with the lean skill costs 1.8 to 3.2 times the tokens of the reference loop in tok (9.7 and 8.6 thousand tokens more per run for Claude Sonnet 5.5, 14.7 and 10.6 thousand for Claude Haiku 5.5, 24.0 and 26.5 thousand for gpt-oss 120B, 22.4 and 17.0 thousand for Beam, $p_\text{Holm}<0.001$ for every model and set), because a loop controlled by the model has at least two turns which both carry the skill and the tool definitions, while the reference loop often ends after one attempt. The validity of the lean agent is equal to or higher than the validity of the reference arm for every model, without a significant difference, and the output of the lean agent exceeds the output of the reference arm by at most 0.5 thousand tokens.

The two BPMN XML agent arms of Beam measure the tool-call channel of the model. Of Beam's calls of \texttt{check\_bpmn}, 94 and 93\,\% arrive with empty arguments under the distributed skill on PET and PMo and 99 and 96\,\% under the lean skill, so a document reaches the checker, under the distributed and the lean skill, in 7 and 5 of the 42 PET runs and in 5 and 8 of the 45 PMo runs, the agent spends its 12 calls on repeated attempts, 47.6 and 24.4\,\% of the final models are valid under the distributed skill and 23.8 and 24.4\,\% under the lean skill, and a run takes 74 to 90 thousand tokens. The probe locates the cause in the name of the parameter. The documents given to \texttt{check} as \texttt{text}, the tok model, the tok text repeated to the length of the XML and the plain text, arrive in 9 of 9 calls, up to 3\,904 characters and 162 double quotes, whereas the documents given to \texttt{check\_bpmn} as \texttt{xml}, the XML export, the XML cut to the length of the tok text and the plain text, arrive in 2 of 9 calls. The XML given to \texttt{check\_bpmn} with its parameter renamed \texttt{text} arrives in 3 of 3 calls, and the XML given to \texttt{check} as \texttt{text} arrives in 3 of 3 calls, while the two variants which keep the name \texttt{xml}, one without the parameter \texttt{description} and one with \texttt{xml} as the only parameter, deliver the XML in 1 and 0 of 3 calls. Beam's tok arms, which pass the document as \texttt{text}, are unaffected.

The fidelity of valid models to the texts is the same in tok and in BPMN XML and the same under the two skills. Flow F1 with invalid models scored 0 lies within 0.08 across the arms of a model, outside the two BPMN XML agent arms of Beam and the arm of gpt-oss 120B with the distributed BPMN skill on PMo, whose invalid models lower the score to 0.260. Among the primary tests, no difference in flow F1 outside the two Beam arms is significant after Holm's adjustment, the largest being Claude Haiku 5.5 as a lean agent against the distributed agent on PET with $0.064$ ($0.018$ to $0.109$, $p_\text{Holm}=0.76$), and the one with the smallest adjusted $p$ value Claude Sonnet 5.5 as a lean agent against the reference arm on PMo with $-0.045$ ($-0.079$ to $-0.016$, $p_\text{Holm}=0.25$). Flow F1 is lower on PMo (0.26--0.43) than on PET (0.42--0.55) in every arm outside the Beam arms in BPMN XML, and the two gold standards differ in kind, since a PMo gold model is one drawing of the text by its author and the PET gold ordering comes from annotations of the text.

\subsection{Process mining}

tok imports all 48 BPMN models of the PMo dataset and simulates 47 models, whereas one imported model carries a checker error (R013, a process name on a pool without a process block) and is left out of the simulation. The 47 event logs hold 500 cases each, with a median of 5\,682 events and 116 trace variants per log.

The simulated logs follow the original models, since 38 of the 47 logs, replayed by PM4Py on the original BPMN files, fit with fitness 1 and every trace replayable. In six further models the original BPMN file holds intermediate events, which leave no event in a log, and with the intermediate events bypassed in the BPMN file the same six logs fit with fitness 1 as well, so that the shortfall lies in the conversion of the events to the Petri net. One model has an event-based gateway whose events carry no wait in the simulation, and 251 of its 500 cases stop at the gateway. Two models join branches with inclusive or complex gateways, which PM4Py converts to Petri nets outside the class that alignments accept, and token-based replay gives fitness 0.96 and 0.92.

tok reads the models that a process-mining tool discovers. The inductive miner discovers a model from each of the 47 logs with fitness 1 and a median precision of 0.947 (mean 0.902, minimum 0.417). tok imports all 47 discovered BPMN files without errors, and all 47 keep the kinds of their nodes and flows through the round trip from BPMN to tok and back. The \texttt{tok check} warnings of 46 discovered models are all W021, a branch of an exclusive split without a label, a condition or a default, since a discovered model carries no branch conditions. The ordering of activities in the discovered model agrees with the original model with a median F1 of 1.0 (mean 0.926, minimum 0.2).

PM4Py 2.7.15 takes sequence flows only from the \texttt{incoming} and \texttt{outgoing} elements of the nodes, which BPMN 2.0.2 makes optional and tok 1.0.3 leaves out, so PM4Py finds zero sequence flows in each of the 47 BPMN files that \texttt{tok bpmn} writes from the simulation models.

\subsection{Performance}

The engine works at interactive speed, in that the released program starts in 15\,ms, the median time of a check, a canonical print, a BPMN export with layout and an SVG drawing of a model of corpus N lies between 20 and 24\,ms with the start included, and the import of a MIWG reference model takes 25\,ms. A generated model of 3\,000 flow nodes is checked in 0.11\,s and exported with layout in 0.43\,s. For generated models of 100 to 3\,000 flow nodes, the fitted exponent of time against size is 0.65 for the check and 0.89 for the export with layout.

\subsection{Conformance}

The tagged sources of release 1.0.3 pass every conformance suite of the project. The engine evaluates the DMN Technology Compatibility Kit with 126 of 126 results at compliance level 2 and 3\,313 of 3\,331 results (99.5\,\%) at compliance level 3. The 22 MIWG reference models and the 16 Camunda Modeler files keep their semantic content through import, printing and export, and the 67 documents of the tok conformance corpus give the expected canonical text, BPMN XML and diagnostics.

\section{Discussion}\label{sec:discussion}

The size of a model decides what a language model pays for writing the model. A tok text takes about half of the tokens of the smallest standard BPMN of a file drawn in a tool and about an eighth of the tokens of the file as exported by the tool, so that a language model which writes tok spends its output on the content of the model. The saving shows in the study as fewer tokens per run for Claude Sonnet 5.5, the model with the shortest runs, whose output in tok is about a third of the output in BPMN XML on both sets of texts, and the direction of the saving agrees with the token reductions reported for compact representations of process models \citep{Agrawal2026bpcf,Kopke2025efficient}.

The capability of a language model decides what the model makes of tok. The four models learn tok from one page of reference and write valid tok for 85.7 to 100\,\% of the texts, in the manner in which a model translates an unseen language from a grammar book placed in its context, with a success that depends on the model \citep{Tanzer2024benchmark}, and writes code for an unseen library from the documentation of the library \citep{Zhou2023docprompting}. The knowledge of BPMN XML that the models bring from their training lets every model write valid BPMN XML for every text. Claude Sonnet 5.5 writes tok as reliably as BPMN XML, with fewer tokens and at about two fifths of the cost of a run in BPMN XML, whereas gpt-oss 120B needs more attempts in tok and writes valid tok for 85.7 and 91.1\,\% of the texts. The agent loop with the lean skill keeps Claude Sonnet 5.5 and Claude Haiku 5.5 at 100\,\% on both sets in both representations and lifts gpt-oss 120B to 97.6 and 97.8\,\% in tok, at 1.8 to 3.2 times the tokens of the reference loop in tok and, for Claude Sonnet 5.5, at 21 and 14\,\% fewer tokens than the same loop in BPMN XML. The design of the skill decides the cost of the loop more than the representation does. The distributed skill of tok sends the model to read the reference through a tool, the text of the reference returns with every later turn of a conversation, whose length costs tokens and, in long contexts, the use of the information in the middle of the context \citep{Liu2024lost}, and the loop costs 22 to 45 thousand tokens more per run than the lean skill at equal or lower validity. A run of Claude Sonnet 5.5 as a lean agent costs \$0.024 and \$0.022 on the two sets, against \$0.069 and \$0.065 for the same agent in BPMN XML and \$0.016 and \$0.019 for a run in tok from the reference.

The checker carries the knowledge of the standard to the writer of a model. A language model that meets tok for the first time learns the syntax from a page of reference, and the checker supplies the rules of BPMN 2.0.2 one diagnostic at a time, each with the line, the element and the clause, in the same form in which the diagnostics guide a student in an editor. The limits of the checker lie in two kinds of fault, an exclusive split without conditions and a deadlock between an exclusive split and a parallel join, which are behavioural properties that a check of single elements reaches only in part and whose complete detection belongs to soundness checkers which explore the state space of a model \citep{Fahland2011analysis}. Grammar-constrained decoding offers an alternative to the loop, in which the output of a model is forced into a context-free grammar at generation time \citep{Geng2023grammar,Poesia2022synchromesh}, whereas the loop of tok leaves the model free and checks the result, so that one channel serves a model in a loop and a person in an editor, and the context conditions beyond the grammar, such as unique identifiers and resolved references, are checked by the same channel.

For the theory of communication between people and language models, the study gives a measured case of grounding through an artefact, in the sense in which \citet{ClarkBrennan1991} describe grounding as the exchange of evidence of understanding. One page of reference and the messages of a checker ground the strongest models in a notation which the models have never seen, at a cost in tokens below the cost of the format which the models know, and the agent loop grounds the same models through a tool at 1.8 to 3.2 times the tokens of the reference loop when the skill carries the reference and at 4.4 to 6.5 times when the skill sends the model to read the reference through a tool. The three conditions are grounding channels with different costs, and the diagnostics which every channel carries are local because the formal structure of the notation is local. The channel itself belongs to the ground, as the probe of Beam shows: the same document reaches the checker in every call when the tool names its parameter \texttt{text} and in few calls when the tool names the parameter \texttt{xml}, a property of the encoding of the call and of neither notation. A property of the grammar, one statement per line with named references, therefore appears in the data as a repair loop which ends after 1.2 to 2.3 attempts on average in the reference arms, the same kind of loop in which a model repairs code from the feedback of an execution \citep{Chen2024selfdebug} or from a verbal signal about its previous attempt \citep{Madaan2023selfrefine,Shinn2023reflexion}. The costs of the two channels belong in the comparison next to the validity, as the evaluation of agents requires \citep{Kapoor2025agents}. For the design of notations the same observation supports a rule, according to which a textual concrete syntax over an existing metamodel keeps every statement local and every reference named, because the checker then localises 99.7\,\% of the faults which the checker reports to the element or the line, and a writer, human or machine, repairs from the localisation.

For research on process modelling with language models, the study offers an instrument and a result. The instrument consists of a representation which the model writes and a person reads, a checker whose diagnostics feed back into the loop, a judge independent of the notation, and a run script which records every prompt, answer, model file and validator output, so that comparisons of representations, whose quality criteria differ from study to study \citep{Brissard2026representation,Agrawal2026bpcf,Lauer2026competences}, gain a common judge and a public baseline with the cost of a run reported next to the validity \citep{Kapoor2025agents}. The result places the representation among the variables with measured effects, next to the model, the skill design and the feedback channel, since the same text costs the strongest model a third of its output, costs a weaker model more attempts, and costs every model 1.7 to 2.8 times more when the skill makes the model read the reference through a tool.

tok sits at both ends of process mining, in the sense that the simulator turns a model into an event log which a process-mining tool replays on the original model, and the importer turns a discovered model into a text which a person reads and a checker verifies. The demonstration also exposes a detail of interchange, since a BPMN file without the optional \texttt{incoming} and \texttt{outgoing} elements is valid, while PM4Py \citep{Berti2023PM4Py} reads the flows of a model only from the two optional elements, one more instance of the differences between tools in the reading of the same format \citep{Geiger2018bpmn}. Writing the optional elements is a small change of the exporter of tok and closes the gap.

A process model in tok behaves as a document, which changes how an organisation keeps process knowledge. The text goes into version control with differences of one line per change, the benefit for which text-based modelling argues \citep{Gronniger2014textbased}, into reviews and reports, and into the prompt of a language model, while the same engine turns the text into the files which modelling tools and process engines read, so that the model, the review of the model and the instruction to a model share one artefact. A model written as text is also open to screen readers and to people who work with text in place of pictures, for whom diagrams are otherwise reached through textual or tactile renderings \citep{King2004presenting,Loitsch2012viable}, a claim which rests on the design of the notation and for which a study with users is still to be made.

The study supports a premise for process management with language models, shared with the vision of large process models \citep{Kampik2025} and with the opportunities named for language models in process management \citep{Vidgof2023llm}, according to which a person and a model can work on the same short text with an engine that checks every change against the standard and compiles the result for the tools of the organisation. The evidence covers the writing of single models from descriptions by four models, and the editing of existing models, the collaboration of people with models over time and the execution of the compiled processes remain open questions.

\section{Limitations}\label{sec:limitations}

The language-model study covers two sets of process descriptions in English, PET and PMo, and four language models, namely two Claude models of Anthropic and the open-weight models gpt-oss 120B and Beam. The gold standards differ between the two sets, because PET annotates activities and flow relations in the text \citep{Bellan2023PET}, whereas PMo provides one BPMN model per description drawn by the authors of each source \citep{Brissard2025PMo}, so that fidelity measures agreement with one reading of each text.

The judge decides validity by the OMG XML Schema \citep{OMG2014BPMN}, the correctness rules of bpmnlint \citep{bpmnlint} and a minimum of structure. A valid model can still misread the text, and fidelity measures the reading only through the names and the ordering of activities.

Hosted models change over time \citep{Chen2024chatgpt} and sample at random. The study records the model identifiers and the dates of the calls, the conclusions rest on paired differences over texts, and the PMo runs and the agent arm have one repetition per text.

The agent conditions differ from the reference condition in several respects at once, which are the skill document, the tools, the loop controlled by the model, and release 1.0.3 of tok with four further warnings, and the two skill designs differ in the length of the skill and in the number of tools at once. The comparisons between conditions therefore measure packages, and the skill effect within tok measures the reading policy of the skill together with the tool set. Beam's calls of \texttt{check\_bpmn} arrive without the document, so the two agent arms of Beam in BPMN XML measure the tool-call channel of one model and carry no information about the representation.

The author of tok wrote the notation, the one-page references, the skill document, the corpus of tok models and the fault operators. The tool files, the reference models, the decision models, both sets of texts, the judge and the reader of the fidelity measurement come from external sources.

The process-mining demonstration fixes durations and branch probabilities by a rule, so that the logs show the behaviour which the models allow without the frequencies of real work, and the demonstration uses one discovery algorithm with one setting.

\section{Conclusion}\label{sec:conclusion}

tok is a plain-text notation for the content of BPMN 2.0.2 and DMN 1.5, with one engine that checks, compiles, imports and simulates models. The notation is an alternative concrete syntax of the standard metamodels, which keeps the translation to the standard exact in both directions, and a symbolic notational system which moves the geometry of a drawing out of the text, and the checker of the notation is a grounding channel whose evidence is local because the grammar is local. The study shows models two to four times smaller than the smallest standard XML, a checker that signals 98\,\% of injected faults, language models that learn tok from one page and write valid models for 85.7 to 100\,\% of the texts with fewer tokens for the strongest model, logs that fit their models, discovered models that tok reads back, and an engine that passes the conformance suites of DMN and of BPMN interchange. A person, a tool and a language model can therefore hold one process in one text, and the price of the shared text is measured for each of them.

\section*{Availability}

tok is free software under GPL-3.0-or-later, archived with release 1.0.3 at Zenodo \citep{Deja2026tok}. The skill documents in the repository of tok and in the agent plugin of tok, version 1.0.3.1, follow the lean skill of the study, in that the model reads the one-page reference once and opens the specification only for a diagnostic the model cannot act on. The tables behind every number of the paper accompany the paper, and the scripts and every prompt, answer, model file and validator output of the study are archived at Zenodo \citep{Deja2026evidence}.

\section*{Use of generative AI}

Generative AI tools (Claude models of Anthropic, used through Claude Code) were used under the author's direction in four roles: resolving software, \LaTeX{} compilation and rendering errors; building and maintaining the scripts that carry the results of the evidence repository into the data files, tables and figures of the manuscript, with every number in the text checked against the data files; curating the deposited repositories and distributing the software and the study materials across repositories and package registries; and copy-editing and restructuring manuscript prose for clarity, terminology consistency and length. The language, the research questions, the study design, the analyses, their interpretation and the writing are the author's. All text and code were reviewed and verified by the author, who takes full responsibility for the content of the article. The language models evaluated in the study are the subject of the study and took no part in the analysis.

\bibliographystyle{plainnat}
\bibliography{references}

\end{document}